\documentclass[preprint,11pt]{JHEP3}% 10pt is ignored!

\JHEPspecialurl{http://jhep.sissa.it/JOURNAL/JHEP3.tar.gz}

\usepackage{epsfig,multicol,amsmath}

\usepackage{bm}  % for the command \bm (boldface greek letters)
\usepackage{amsmath}     % defines the command \dfrac (fractions in displaystyle
\usepackage{epsfig,epsf}
\usepackage{graphicx}
\usepackage{rotating}
\usepackage{cite}

\def\beq{\begin{equation}}
\def\eeq{\end{equation}}
\def\bea{\begin{eqnarray}}
\def\eea{\end{eqnarray}}

\def\eq#1{{Eq.~(\ref{#1})}}
\def\fig#1{{Fig.~\ref{#1}}}

\newcommand{\as}{\alpha_S}

\newcommand{\Lb}{\left(}
\newcommand{\Rb}{\right)}
\parskip=\itemsep               %?

\newcommand{\nn}{\nonumber}

\newcommand{\h}{\frac{1}{2}}

\newcommand{\pom}{I\!\!P}

\def\pom{{I\!\!P}}

\title{Description of  LHC data in a  soft interaction model.}
\author{\Large  E. Gotsman$^{a}$\thanks{Email:
gotsman@post.tau.ac.il.}\,, E. Levin$^{a,b}$\thanks{Email:
leving@post.tau.ac.il}\,\,and\,\,U. Maor$^{a}$\thanks{Email: maor@post.tau.ac.il.}\, 
\\
a)\,  \,Department of Particle Physics, School of Physics and Astronomy,
Raymond and Beverly Sackler
 Faculty of Exact Science, Tel Aviv University, Tel Aviv, 69978, Israel\\ 
b)\,\,Departamento de F\'\i sica, Universidad T\'ecnica Federico Santa 
Mar\'\i a, Avda. Espa\~na 1680\\ and Centro 
Cientifico-Tecnol$\acute{o}$gico de Valpara\'\i so,Casilla 110-V, 
Valparaiso, Chile

\\}

\abstract{We show in this paper that we have found 
 a set of parameters in our model for the soft interactions 
at high energy, that successfully describes all high energy experimental 
data,   including the LHC data.
 This model is based on a
 single Pomeron with large intercept
 $\Delta_\pom = 0.23$ and slope $\alpha'_\pom = 0$,
 that describes both long and short distance processes. It also provides
 a natural matching with perturbative QCD. All features of our
 model are similar to the expectations of N=4 SYM, which at present is the 
only theory
 that is able to treat  srong interactions on a theoretical basis.
}

\dedicated{PACS: 13.85.-t, 13.85.Hd, 11.55.-m, 11.55.Bq}
\preprint{TAUP -2155/12\\
{\tt }\\
\today}
\begin{document}

\section{Introduction}

It is well known that a comprehensive theory of strong interactions is 
still in
an embryonic stage, as we understand
little about non-perturbative QCD. On the other hand, considerable
progress has been
made  in building models for soft scattering at high 
energies\cite{GLM1,GLM2,KAP,KMR,OST}. The main 
constituent of these 
models is the soft Pomeron with a relatively large  intercept 
$\Delta_{\pom} = \alpha_{\pom} - 1 = 0.2  - 0.4$, 
and exceedingly small slope 
$\alpha_{\pom}^{\prime} \simeq 0.02\,GeV^{-2}$.
A Pomeron with these characteristics appears in N=4 SYM
\cite{BST,HIM,COCO,BEPI,LMKS} with a large coupling, which, 
at the moment, is the only theory that allows us
to treat the strong interaction on a theoretical basis. Having
$\alpha_{\pom}^{\prime} \to 0$, the Pomeron in 
these models has a natural matching with the hard Pomeron that
evolves from perturbative QCD. Hence,
these models could be a first step in building a
self-consistent theoretical description of 
soft interactions at high energy, 
in spite, of the many phenomenological parameters 
(of the order of 10-15) appearing in these models.

The new LHC data on soft interaction scattering at high energy 
(see Refs.\cite{ALICE,ATLAS,CMS,TOTEM}) shows that this hope was premature.
Indeed, all of the models
\cite{GLM1,GLM2,KAP,KMR,OST} failed to describe the LHC data.
At first sight, this 
suggests the need for a comprehensive 
revision of the main ingredients of the Pomeron models 
applied to high energy soft interactions.

However, an alternate approach is to  adjust the Pomeron 
model's parameters.
Our first attempt to do this was published in Ref.\cite{GLMLAST}, 
In which we concluded: 
"In spite of the fact that the values of the parameters,   
extracted from our current fitting, are slightly different from our 
previous values, the overall picture remains unchanged. 
Our updated total and elastic cross sections
are slightly lower than the published TOTEM values\cite{TOTEM}, but still 
within the relatively large experimental error bars. Should future LHC 
measurements confirm the present TOTEM values, 
we will need to revise our dynamic picture for soft scattering."

In the present paper we retract the above statement, 
as we find that the  above conclusion  was premature. 
 The set of parameters in our previous
paper was found by fitting all data with energy $W \geq 500 GeV$,
including the LHC data. 
In the present version of our GLM model we made no changes other than 
tuning the 3 Pomeron parameters.  
Our tuned $\Delta_{\pom}$ changes from 0.21 to 0.23, while $G_{3\pom}$
and $\gamma$, the Pomeron-proton vertices, are unchanged. The small change 
in the value of $\Delta_{\pom}$ is sufficient to produce the desired 
results in our $\sigma_{tot}$ and $\sigma_{el}$ output values
for LHC energies, while 
the changes in the output values of the other observables are small enough 
not to spoil the good reproduction of the data achieved in 
Ref.\cite{GLMLAST}.

Relying on our model that describes all available  data, we also include 
in this 
paper  predictions for LHC 
higher energies ($W = 8 \,TeV$ and $W = 14\, TeV$).
%THE ENERGY OF AN AUGER EVENT WHICH ATTRACTED SOME RECENT 
%INTEREST (SEE REF.\cite{GLMLAST}).
As there has been a recent renewal of 
interest in cross sections 
eminating from cosmic ray data \cite{BLHA}, we also list our results for 
$W = 57\, TeV$.

\section{Our Model}
To make our presentation self-contained we give a brief overview of 
our model, in spite of the fact, 
that all the ingredients and formulae have been published (see 
Ref.\cite{GLM1,GLM2, GLMLAST}).
%%%%%%%%%%%%%%%%%%%%%%%%%%%%%%%%%%%%%%%%%%%%%%%%%%%%%%%%%%%%%%%%%%%%%%%%%%%%%
\subsection{Two channel model}
 
To account for diffraction dissociation in the states with masses  
that are much smaller than
the initial energy, we use the simple two channel 
Good-Walker model. In this model we 
introduce two eigenwave functions, 
$\psi_1$ and $\psi_2$, which
diagonalize the 2x2 interaction matrix ${\bf T}$,
\beq \label{2CHM}
A_{i,k}=<\psi_i\,\psi_k|\mathbf{T}|\psi_{i'}\,\psi_{k'}>=
A_{i,k}\,\delta_{i,i'}\,\delta_{k,k'}.
\eeq

The two observed states are an hadron whose wave function we denote
by $\psi_h$,
and a diffractive state with a wave function $\psi_D$. These two 
observed states can be written in the form
\beq \label{2CHM31}
\psi_h=\alpha\,\psi_1+\beta\,\psi_2\,,\,\,\,\,\,\,\,\,\,
\psi_D=-\beta\,\psi_1+\alpha \,\psi_2\,,
\eeq
where, $\alpha^2+\beta^2=1$
%%%%%%%%%%%%%%%%%%%%%%%%%%%%%%%%%%%%%%%%%%%%%%%%%%%%%%%%%%%%%%%%%%%%%%%%%%%%%%

\subsection{ Eikonal approach}

Using \eq{2CHM}, we can rewrite the s-channel unitarity constraints in the 
form
\beq \label{UNIT}
2\,\mbox{Im}\,A_{i,k}\left(s,b\right)=|A_{i,k}\left(s,b\right)|^2
+G^{in}_{i,k}(s,b),
\eeq
where, $G^{in}_{i,k}$ denotes the contribution of all
non GW inelastic processes.

In a general solution of \eq{UNIT}
\beq \label{2CHM1} 
A_{i,k}(s,b)=i \Lb 1
-\exp\Lb - \frac{\Omega_{i,k}(s,b)}{2}\Rb\Rb, \eeq \beq \label{2CHM2}
G^{in}_{i,k}(s,b)=1-\exp\Lb - \Omega_{i,k}(s,b)\Rb,
\eeq
in which $\Omega_{i,k}$  are arbitrary.
In the eikonal approximation $\Omega_{i,k}$
are taken as being real.  In our model we choose  $\Omega_{i,k}$ as
the contribution of a single Pomeron exchange (see \eq{Born}).

From \eq{2CHM2} we deduce that the
probability that the initial state $(i,k)$ remains intact
during the interaction, is $P^S_{i,k}=\exp \Lb - \Omega_{i,k}(s,b) \Rb$. 

It should be stress that in this model we describe the variety of
states produced  with different masses and other quantum numbers, 
in diffraction dissociation, by a single state
($\psi_D$) with unknown mass. 
Using \eq{2CHM},\eq{2CHM31} and \eq{Born} we obtain 
that one Pomeron exchage has the following
contributions to the different processes:\\
%%%%%%%%%%%%%%%%%%%%%%%%%%%%%%%%%%%%%%%%%%%%%%%%%%%%%%%%%%%%%%%%%%%%%%%
\bea \label{POMDF}
\mbox{Elastic\,\, scattering:}\,\,p + p \to p + p &
~~~~~~~~~~& A_{p,p}\,=\, \alpha^4 A_{1,1}\,\,+\,\,2 \alpha^2\,
\beta^2 A_{1,2}\,\,+\,\,\beta^4 A_{2,2};\nn\\
\mbox{Single\,\, diffraction:}\,\,p + p \to D + p  & 
~~~~~~~~~~~&A_{p, D}  
\,=\,\alpha \beta\Lb - \alpha^2 A_{1,1} + ( \alpha^2 - \beta^2) 
 A_{1,2}\,+
\,\beta^2 A_{2,2}\Rb;\nn\\
\mbox{Double\,\,  diffraction:}\,\,p + p \to D + D&
~~~~~~~~~~&A_{D, D}\,=\,\alpha^2 \beta^2 \Lb A_{1,1}\,
-\,2\,A_{1,2}\,+\,A_{22}\Rb;
\eea

\eq{POMDF} shows the particular realization of the Good-Walker 
 mechanism\cite{GW}  for diffraction production, in the two channel model.
 It is worthwhile mentioning that this mechanism is
 the only source of diffraction in N=4 SYM, due to AdS-CFT correspondence
 (see  for example Ref.\cite{LMKS}).

\subsection{ Pomeron}

As we have mentioned, our basic ingredient is the soft Pomeron,
 which has the following contribution to
 the elastic scattering amplitude for the process:
 $i + k \to i + k$ where $i$ and $k$ are the states that 
are described by the wave functions $\psi_i$ and $\psi_k$  
\beq \label{Born}
\Omega_{i,k}(s,b) \,\,=\,\, g_i(b)\,g_k(b)\,P(s). 
\eeq
$P(s)\,=\,s^\Delta$, and $g_i(b)$ and $g_k(b)$ are the 
Pomeron-hadron vertices parameterized in the form: 
\beq \label{GP} g_l\Lb 
b\Rb\,=\,g_l\,S_l(b)\,=\,\frac{g_l}{4 \pi}\,m^3_l \,b\,K_1\Lb m_l b\Rb. 
\eeq 
$S_l(b)$ is the Fourier transform of $\frac{1}{(1 + q^2/m^2_l)^2}$, 
where, $q$ is the transverse momentum carried by the Pomeron,  
L=I,K.

The form of $P(s)$ that we use, corresponds to  
a Pomeron trajectory slope $\alpha'_\pom = 0$. 
This is compatible with the exceedingly small fitted value 
of $\alpha_{\pom}^{\prime}$, and in accord with N=4 SYM \cite{BST}.

$\Omega_{i, k}$ is the imaginary part of the scattering amplitude 
for a single Pomeron exchange.

\subsection{Pomeron interactions}
In the case of $\Delta_{\pom} \to 0$, we know that the Pomeron interaction
leads to a new source of diffraction production with large mass
($M \propto s$), which cannot be described by the Good-Walker mechanism.
For $\Delta_{\pom} > 0$, the Pomeron interaction is responsible for 
diffraction production of finite (independent of the total energy) mass:
$\log\Lb M^2/s_0\Rb \propto 1/\Delta_{\pom}$. This source can be
treated in the same manner as the Good-Walker mechanism 
(see Ref.\cite{GU}).  
 We believe
that we need to account for the Pomeron interaction separately, since the
typical diffractive mass depends on the Pomeron intercept, and this 
effect needs to be taken into account in the fitting procedure.

 Taking $\alpha'_\pom=0$, allows us to sum all diagrams
having   Pomeron interactions\cite{GLM1,GLM2}. This is the great 
advantage
of such an approach. In our model we only take into account  triple 
Pomeron
interaction vertices ($G_{3\pom}$), this provides a natural matching to 
the hard Pomeron, since  at short distances $G_{3\pom} \propto \as^2$, 
while other vertices are much smaller.
For a thorough description of the procedure for summing all diagrams, we 
refer to our papers (see Refs.\cite{GLM1,GLM2,GLMLAST}.

{\it Enhanced diagrams:} 
In our model\cite{GLM2}, the Pomeron's Green function that
includes all enhanced diagrams, is approximated using the MPSI 
procedure\cite{MPSI}, in which a multi Pomeron interaction
(taking into account only triple Pomeron vertices) is          
approximated by large Pomeron loops of rapidity size of $\ln s$.

We obtain 
\beq \label{GFPEN}
G_{\pom}\Lb Y\Rb\,\,=\,\,1 \,-\,\\exp\Lb \frac{1}{T\Lb Y\Rb}\Rb\,
\frac{1}{T\Lb Y\Rb}\,\Gamma\Lb 0,\frac{1}{T\Lb Y\Rb}\Rb, 
\eeq
in which:
\beq \label{ES11}
T\Lb Y \Rb\,\,\,=\,\,\gamma\,e^{\Delta_{\pom} Y}.
\eeq
$\Gamma\Lb 0, 1/T\Rb$ is the incomplete gamma function
(see formulae {\bf 8.35} in Ref.\cite{RY}).

{\it Semi-enhanced (net) diagrams:} 
A brief glance at the values of
the parameters  of our model (see Ref.\cite{GLMLAST}), shows that we
have a new small parameter, namely, $G^2_{3 \pom} \,P(s)  \,\ll\,1$,
while $G_{3\pom} \,g_i \,P(s) \,\approx\, 1$.  
We call the diagrams which are
proportional to $\Big(G_{3\pom} \,g_i \,P(s) \Big)^n$, but do not
contain any of $G^2_{3 \pom} \,P(s)$ contributions, net diagrams.
Summing the net diagrams \cite{GLM1}, we obtain 
the following expression for $\Omega_{i,k}(s,b)$:
\beq \label{FIMF}
\Omega^{i,k}_{\pom}\Lb Y; b\Rb\,\,\,= \,\,\, \int d^2 b'\,
\,\,\,\frac{ g_i\Lb\vec {b}'\Rb\,g_k\Lb\vec{b} - \vec{b}'\Rb
\,\Big( 1/\gamma\, G_{\pom}\Lb T(Y)\Rb\Big)}
{1\,+\,\Lb G_{3\pom}/\gamma\Rb G_{\pom}\Big(T(Y)\Big)\,\left[
g_i\Lb\vec{b}'\Rb + g_k\Lb\vec{b} - \vec{b}'\Rb\right]}.
\eeq
$G_{3\pom}$ is the triple Pomeron vertex, 
and $\gamma^2 = \int \frac{d^2 k_t}{4 \pi^2} G^2_{3 \pom}$.

\subsection{Final formulae}
For the elastic amplitude we have:
\beq \label{ES}
a_{el}(b)\,=\,i \Lb \alpha^4 A_{1,1}\,
+\,2 \alpha^2\,\beta^2\,A_{1,2}\,+\,\beta^4 A_{2,2}\Rb. 
\eeq 
For diffraction production we introduce an additional contribution due 
to the Pomeron interaction, which we call non-Good-Walker (non-GW). 
For single diffraction we have (see Ref.\cite{GLMLAST}):
\bea \label{SD}
A^{sd}_{i; k, l}\,\,\,&r =&\,\,\int d^2 b'\,2\,\Delta\,\,
\Big(\frac{G_{3\pom}}{\gamma} \,
\frac{1}{\gamma^2}\Big)\,g_i\Lb \vec{b} - 
\vec{b}',m_i\Rb\,\,g_l\Lb  
\vec{b}',m_l\Rb\,g_k\Lb  \vec{b}',m_k\Rb \nn\\
&\times &\,\,Q\Lb g_i,m_i,\vec{b} -                  
\vec{b},Y_m\Rb\,\,Q\Lb g_k,m_k, \vec{b}',Y - 
Y_m\Rb\,\,Q\Lb g_l,m_l, \vec{b}',Y - Y_m\Rb,
\eea
where,
\beq \label{SD1}
Q\Lb g, m, b; Y\Rb \,\,=\,\,\frac{G_{\pom}\Lb Y\Rb}{ 1\,\,
+\,\,\Lb G_{3\pom}/\gamma\Rb\,g\,G_{\pom}\Lb Y\Rb \,S\Lb b, m \Rb}.
\eeq
For double diffraction we have:
\bea \label{DD}r 
\tilde{A}^{dd}_{i,k}\,\,&=&\,\,\int d^2 b'\,  4 \,\Lb \vec{b} - 
\vec{b}',m_i\Rb\,\,g_k\Lb  \vec{b}',m_k\Rb\nn\\
&\times &\, \,Q\Lb g_i,m_i,\vec{b} - \vec{b},Y - 
Y_1\Rb\,e^{2 \Delta\,\delta Y} \,Q\Lb g_k,m_k, \vec{b}', Y_1 - \delta Y\Rb.
\eea

\eq{SD} and \eq{DD} are the simplifications of the exact formulae of 
Ref.\cite{GLM1}. We have 
checked that they approach the values of the exact formulae reasonably 
well, within $5 - 10\%$.
  
For single diffraction,  
$ Y = \ln\Lb M^2/s_0\Rb$, where, $M$ is the SD mass. 
For double diffraction, 
$ Y - Y_1 = \ln\Lb M^2_1/s_0\Rb$ 
and $ Y_1 - \delta Y = \ln \Lb M^2_2/s_0  \Rb$, 
where $M_1$ and $M_2$ are  the masses of the two bunches of hadrons 
produced in double diffraction.
$s_0 $ is the minimal produced mass, which is about $1 \,GeV$.
The integrated cross section of the SD channel  
is written as a sum of two terms:
the GW term, which is equal to
\beq \label{FSSDGW}
\sigma^{GW}_{sd}\,\,=\,\,\int\,\,d^2b\,
\Lb \alpha\beta\{-\alpha a^2\,A_{1,1}           
+(\alpha^2-\beta^2)\,A_{1,2}+\beta^2 \,A_{2,2}\}\,\Rb^2.
\eeq
The second term describes diffraction production due to non-GW  
mechanism:
\bea\label{FSSDLM}
&&\sigma^{\mbox{nGW}}_{sd}\,\,=\,\,\,\,2 \int d Y_m
\int d^2 b \,\\
&&\left\{\,\alpha^6\,A^{sd}_{1;1,1}\,
e^{- \Omega_{1,1}\Lb Y;b\Rb}\,\,+\,\alpha^2\beta^4 A^{sd}_{1;2,2}\,
e^{- \Omega_{1,2}\Lb Y;b\Rb} + 2\,\alpha^4\,\beta^2 \,A^{sd}_{1;1,2}\,
e^{- \h\Lb \Omega_{1,1}\Lb Y;b\Rb
+ \Omega_{1,2}\Lb Y; b \Rb\Rb} \right. \nn\\
&&\left.\,\,+\,\,\beta^2\,\alpha^4 \,A^{sd}_{2;1,1}\,
e^{- \Omega_{1,2}\Lb Y;b\Rb}\,\,+\,\,2\,\beta^4\alpha^2
\,A^{sd}_{2;1,2}\,e^{- \h\Lb \Omega_{1,2}\Lb Y;b\Rb
+ \Omega_{2,2}\Lb Y; b \Rb\Rb}\,\,+\,\,\beta^6\,\,
A^{sd}_{2; 2,2}\,e^{- \Omega_{2,2}\Lb Y;b\Rb} \right\}. \nonumber
\eea
                 
The cross section of the double diffractive
production is also a sum of the GW contribution, 
\beq \label{FSDD3}
\sigma^{GW}_{dd}\,\,=\,\,\int \,d^2 b \,\, \alpha^2\,\beta^2
\left\{ A_{1,1}\, -\,2\,A_{1,2}\,+\, A_{2,2} \right\}^2,
\eeq
to which we add the term which is 
determined by the non GW contribution, 
\beq \label{FSDD4}
\sigma^{\mbox{nGW}}_{dd}\,\,=\,\,\int\,d^2 b\,
\left\{\alpha^4\,A^{dd}_{1,1}\,e^{- \Omega_{1,1}
\Lb Y;b\Rb}\,+ \,2 \alpha^2\,\beta^2 A^{dd}_{1,2}\,
e^{- \Omega_{1,2}\Lb Y;b\Rb}\,+\,\,\beta^4\,A^{dd}_{2,2}\,
e^{- \Omega_{2,2}\Lb Y;b\Rb}\,\right\}.
\eeq
In our model, the GW sector contributes to both low and high 
diffracted mass, while the non-GW sector contributes only to 
high mass diffraction ($\log\Lb M^2/s_0\Rb \approx\,1/\Delta_{\pom}$).

\section{Results}
The output of our current model is presented in Fig. 1, in which we 
display 
our calculated cross sections and forward elastic slope. It is 
interesting to compare the quality of our present results with  
 our previous output\cite{GLMLAST}. Recall that the two fits have 
almost identical values for the 
free   parameters, with a single change of $\Delta_{\pom}$ 
from 0.21 to 0.23.\\ 
We list  the main features of our results:
\begin{itemize}
\item 
The main feature of our present calculation 
is  the excellent 
reproduction of TOTEM's values for $\sigma_{tot}$ and $\sigma_{el}$. 
The quality of our good fit to $B_{el}$ is maintained.  
As regards $\sigma_{inel}$, our results are in accord with the higher 
values obtained by ALICE\cite{ALICE} and TOTEM\cite{TOTEM};
 ATLAS\cite{ATLAS} and CMS \cite{CMS} quote lower values with 
large extrapolation errors. We refer the reader to \cite{JPR} who suggests 
that the lower values found by ATLAS and CMS maybe due to the simplified 
Monte Carlo that they used to estimate their diffractive background.

\item 
The quality of our output at lower energies,  
when compared with ISR data, 
is not as good as our 
previous results\cite{KAP}, but still acceptable. Recall that Reggeon 
exchange, which is included in our model, plays an 
important role
at the low energy end of our data, and a negligible role at higher 
energies. As our main goal is to provide a good description of the LHC 
data,  we have not tuned the Reggeon parameters, which could lead to an
improved characterization of the ISR measurements.

\item
An interesting observation is that our updated output strongly supports 
the CDF total and elastic cross sections rather than the E710 values.
This is a common feature of other models\cite{DL,KMR}, that have succeded 
in 
reproducing the TOTEM results by making  a radical change in their 
modellings. 
\item 
Note that our model is the only one which offers a good
reproduction of ISR diffraction, and a reproduction 
of the diffractive cross sections at higher energies.

\item
 Our reproduction of SD and DD cross sections is complicated by the   
lack of  common definitions of
signatures and mass bounds on the diffractive components.
All
 models on the market  have  introduced at least two different
mechanisms to describe diffraction production. In our model these
 two mechanisms  are : the Good-Walker production of the diffraction state
 with finite unspecified mass
 (independent of energy and of the values of the parameters
in our
 model); and  the diffraction due to Pomeron
interactions where the typical mass depends on $\Delta_\pom$, and which
 we call non Good-Walker mechanism.
 In other models (see Refs.
 \cite{KAP,KMR,OST}) the two different mechanisms are called:  low mass 
 diffraction and  high mass diffraction. It is assumed that
 the diffractive production of  mass less than
 $M_0 \approx 2 - 3\,GeV$ stems from Good-Walker mechanism,  
 while the production of  masses larger than $M_0$ is
 due to  the Pomeron interaction. This partition appears
 natural in the case where $\Delta_\pom \to 0$, when the Pomeron
 interaction is responsible for  diffractive production
 in the region where the mass  increases with  energy.
However,  for $\Delta_\pom \neq 0$ both mechanisms lead to the production 
of final states with
 typical masses,  and therefore the artificial requirement that the 
Good-Walker
 mechanism produces masses less than $M_0$, appears unnatural.
 In our approach we do not specify the value of the mass in
 the Good-Walker mechanism, but it is certainly larger than $M_0
 = 2-3 \,GeV$. As the Pomeron interaction gives a negligible
 contribution at $W = 900 \,GeV$,
we  estimate that our typical mass for the GW mechanism is about 
$10-20\,GeV$.
 Such a large mass for the  GW contribution, explains why our Pomeron
interaction leads
 to a small contribution to the cross section for
  diffraction production due to Pomeron interactions.
 Note that the experimental errors
on the diffractive cross sections are relatively large,
 and thus the data does not
provide a good enough resolution to differentiate between  competing
models.

%%%%%%%%%%%%%%%%%%%%%%%%%%%%%%%%%%%%%%%%%%%%%%%%%%%%%%%%%%%%%%%%%%%%%%
\item 
Our calculated $d\sigma_{el}(t\leq\,0.55GeV^2)/dt$ is presented in Fig.2 
together with the corresponding data. The quality of the 
fit is very good in this region of t.
 It shows that the impact parameter dependence of the model does
 not change considerably and reproduces the experimental data 
as well, as in our previous version of the model.
  We refrain from trying to reproduce the 
diffractive dip and higher $t$ cross sections since our model is confined 
to the forward cone.
\item
Table 1 summarizes our calculated cross sections and $B_{el}$ at 
1.8 - 57\,TeV. The table provides the predicted values of the 
cross sections 
and forward slope at 8, 14 and  57  TeV. Note that the table indicates 
a  slow approach of our predicted elastic amplitude to 
saturation. A common signature for saturation is 
$\sigma_{el}\,=\,\sigma_{inel}\,=\,0.5\,\sigma_{tot}$.
Considering the energy behaviour of $\sigma_{inel}/\sigma_{tot}$, 
the values of this ratio given by our model suggest 
 a very slow increase to saturation, with 
$\sigma_{inel}/\sigma_{tot}$ = 0.77 at  Tevatron energies 
 decreasing to 0.73 at 57 TeV. \\
 The origin of such a slow approach to 
the black disc limit, turns out to be the same as
 with our previous set of parameters (see Ref.\cite{GLMLAST}):
 where one of the partial scattering 
 amplitude, $A_{1,1}\Lb s, b \Rb   <\,1$ 
at all energies higher than $ 1.8 \,TeV$, 
while $A_{1,2}\Lb s,b\Rb \approx A_{2,2}\Lb s,b \Rb =1$ for $b=0$ at 
the same energies.

\item
 We note that our new set of parameters with
 $\Delta_\pom = 0.23$ does not change any qualitative features
 of the scattering amplitude.
In particular, the impact parameter dependence in our model is   
 similar to the $b$ dependence of our previous model \cite{GLMLAST}.
 Therefore, it is not surprising
that we obtain  good agreement with the experimental
 data for $d\sigma_{el}/d t $ as a function of $t$, at
 different energies (see \fig{dsdt}).
\item
The increase of $\Delta_\pom$ leads to stronger shadowing corrections,
 that manifest themselves in  smaller values for the survival
probabilities
 for the large rapidity gap
($S^2$),
 compared to what we previously obtained with $\Delta_{\pom}$ =0.21
\cite{GLMLAST}. The $S^2$ calculated with our present parameters are:
\bea \label{S2}
&S^2\,=\,7.02\%\,[9.76\%]\,\, at\, W\, =\, 1.8 \,TeV;\,
\,\,\,\,\,\,S^2\,=\,2.98\%\,[6.3\%]\,\, at \, W \,=\,
 7 \,TeV;&\,\,\,\,\nn\\
&S^2\,=\,2.7\%\,[5.9\%]\, \,\,at\, W\, =\, 8 \,TeV;\,
\,\,\,S^2\,=\,1.75\% \,[3.66\%]\,\,at \, W  =\, 14 \,TeV;&
\eea
$[ \dots ]$ denotes the values obtained using our previous set of
parameters (see Ref.\cite{GLMLAST}).

\end{itemize}

\TABLE{\small
\begin{tabular}{|l|l|l|l|l|l|}
\hline
W = $\sqrt{s}$  \,\, TeV
& 1.8
&  7
& 8
&  14
& 57
\\
\hline
 $\sigma_{tot}$ mb    
&  79.2 
& 98.6
& 101
&  109
&130
  \\
\hline
$\sigma_{el}$ mb
& 18.5
&24.6
&  25.2
& 27.9
&34.8
  \\
\hline
$ \sigma_{sd}( M^2 < 0.05s)mb $
&8.2 + [2.07] 
&10.7 + [4.18] 
& 10.9 +  [4.3] 
&  11.5 + [5.81]
&  13 + [8.68]  
 \\
\hline
$\sigma_{dd}$ mb
& 5.12 + [0.38]  
& 6.2 + [1.24]
& 6.32 + [1.29]
& 6.78  + [1.59]
& 7.95 + [5.19]
   \\
\hline
$B_{el}\;GeV^{-2}$
&  17.4
&20.2
& 20.4
&21.6
& 24.6
  \\
\hline
$\sigma_{inel}$ mb 
& 60.7
& 74
& 75.6
&  81.1
&95.2

 \\
 \hline
 $\frac{d \sigma}{d t}|_{t=0}\,mb/GeV^2$
 &326.34
 & 506.4
 & 530.7
 & 608.11
 & 879.2
 \\
\hline \hline 
\end{tabular}
\caption{ Predictions of our model for different energies $W$. The 
quantities appearing in square brackets
 $[ \dots ] $ denote the contribution due to Pomeron interactions
 (so called non-GW part of the cross section)}
\label{t3}}

%%%%%%%%%%%%%%%%%%%%%%%%%%%%%%%%%%%%%%%%%%%%%%%%%%%%%%%%%%%%%%%%%%%%%%%%%%%%%%%%%%%

%%%%%%%%%%%%%%%%%%%%%%%%%%%%%%%%%%%%%%%%%%%%%%%%%%%%%%%%%

\FIGURE[h]{
\begin{tabular}{c c}
\epsfig{file=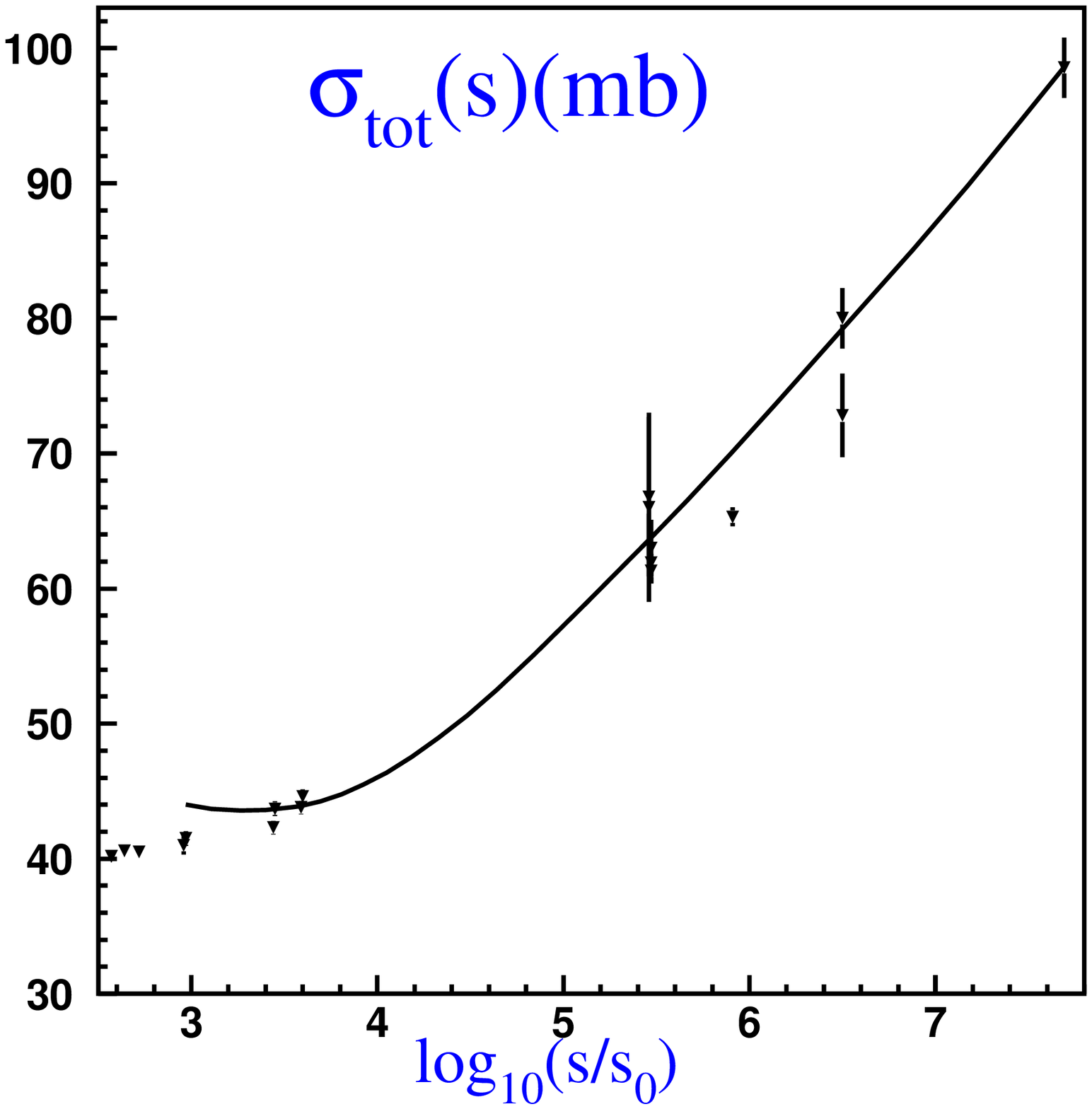,height=50mm,width=65mm}
 &\epsfig{file=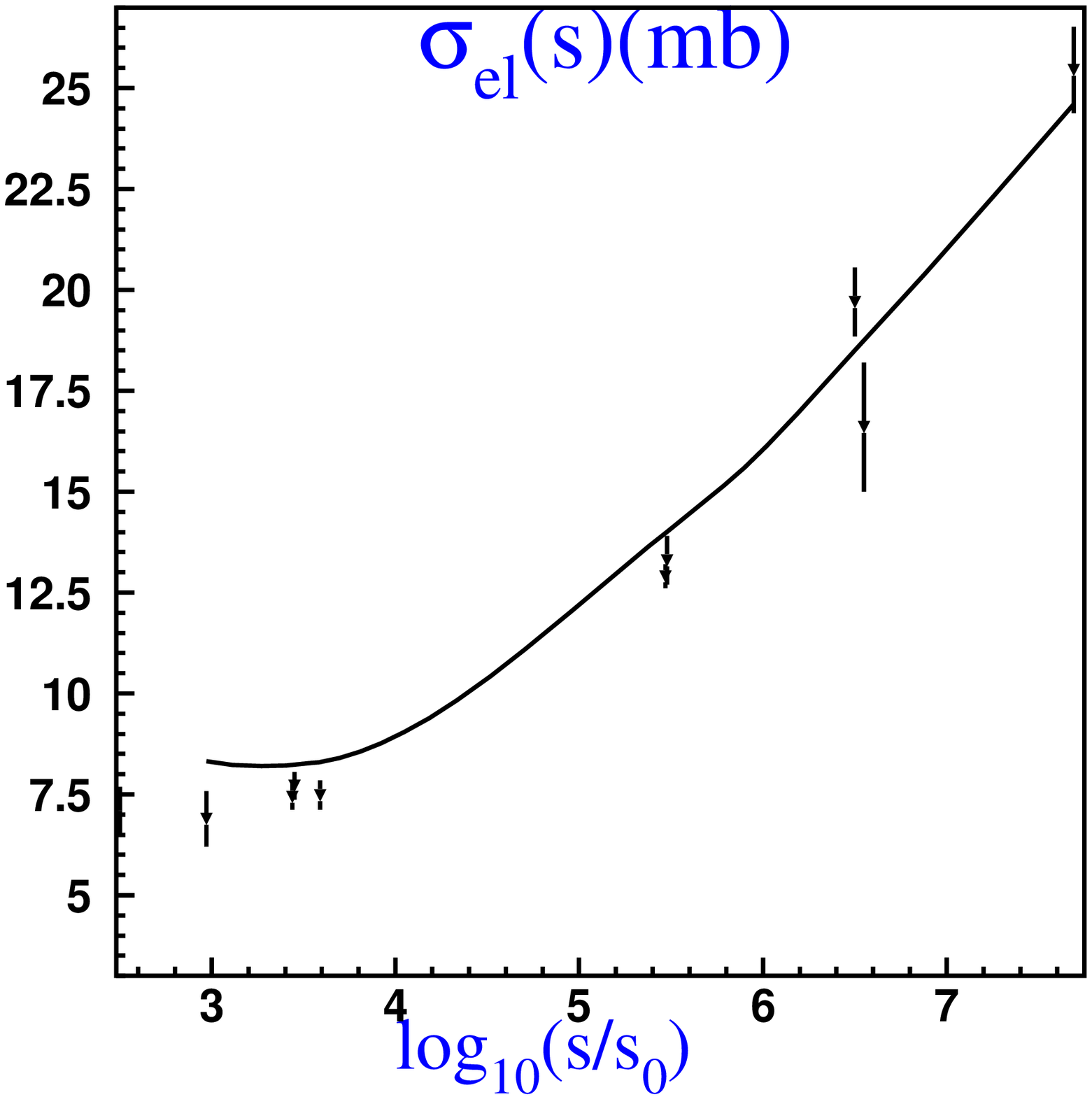,height=50mm,width=65mm}\\
\fig{fit}-a & \fig{fit}-b\\
\epsfig{file=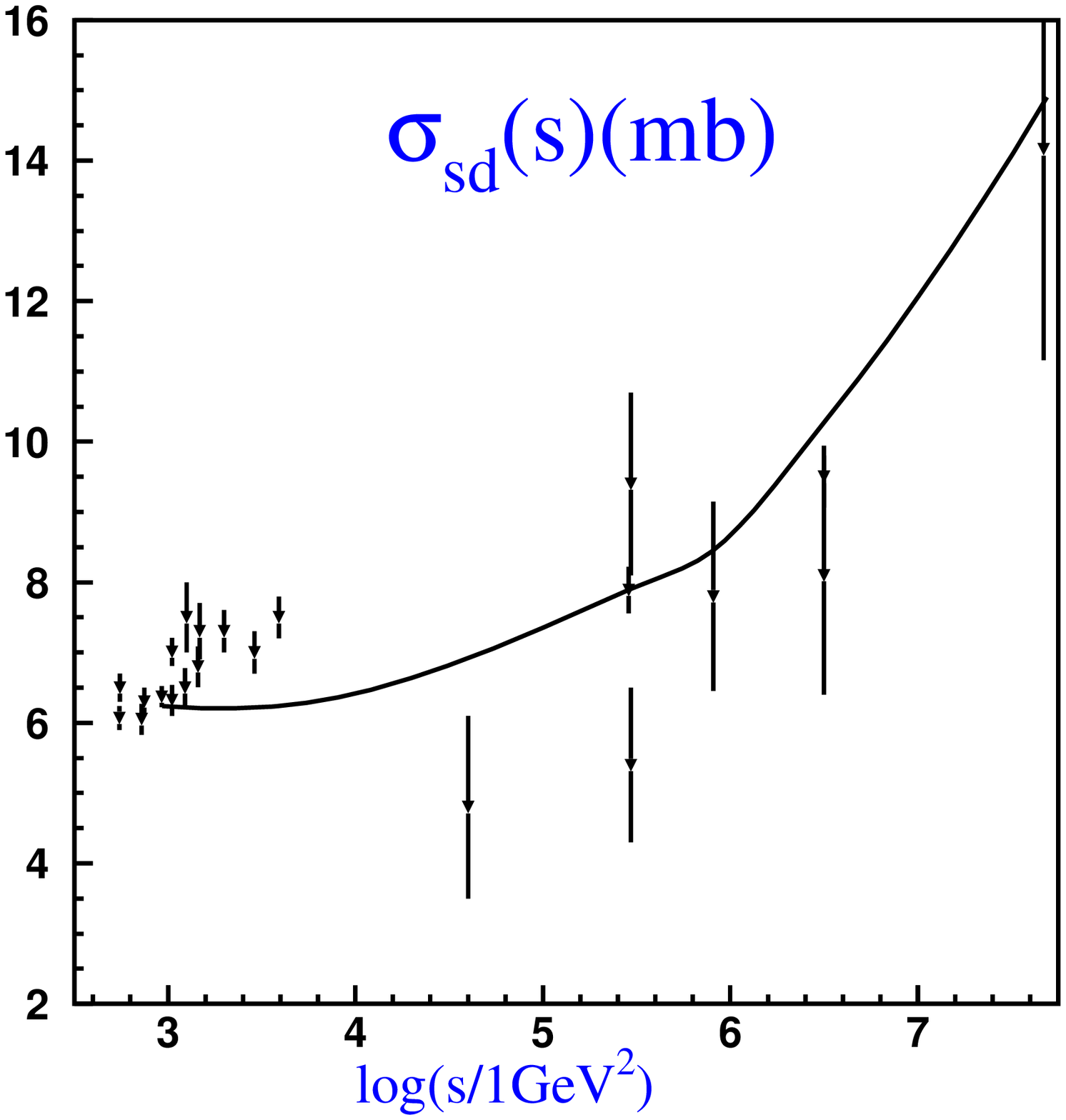,height=50mm,width=65mm}
 &\epsfig{file=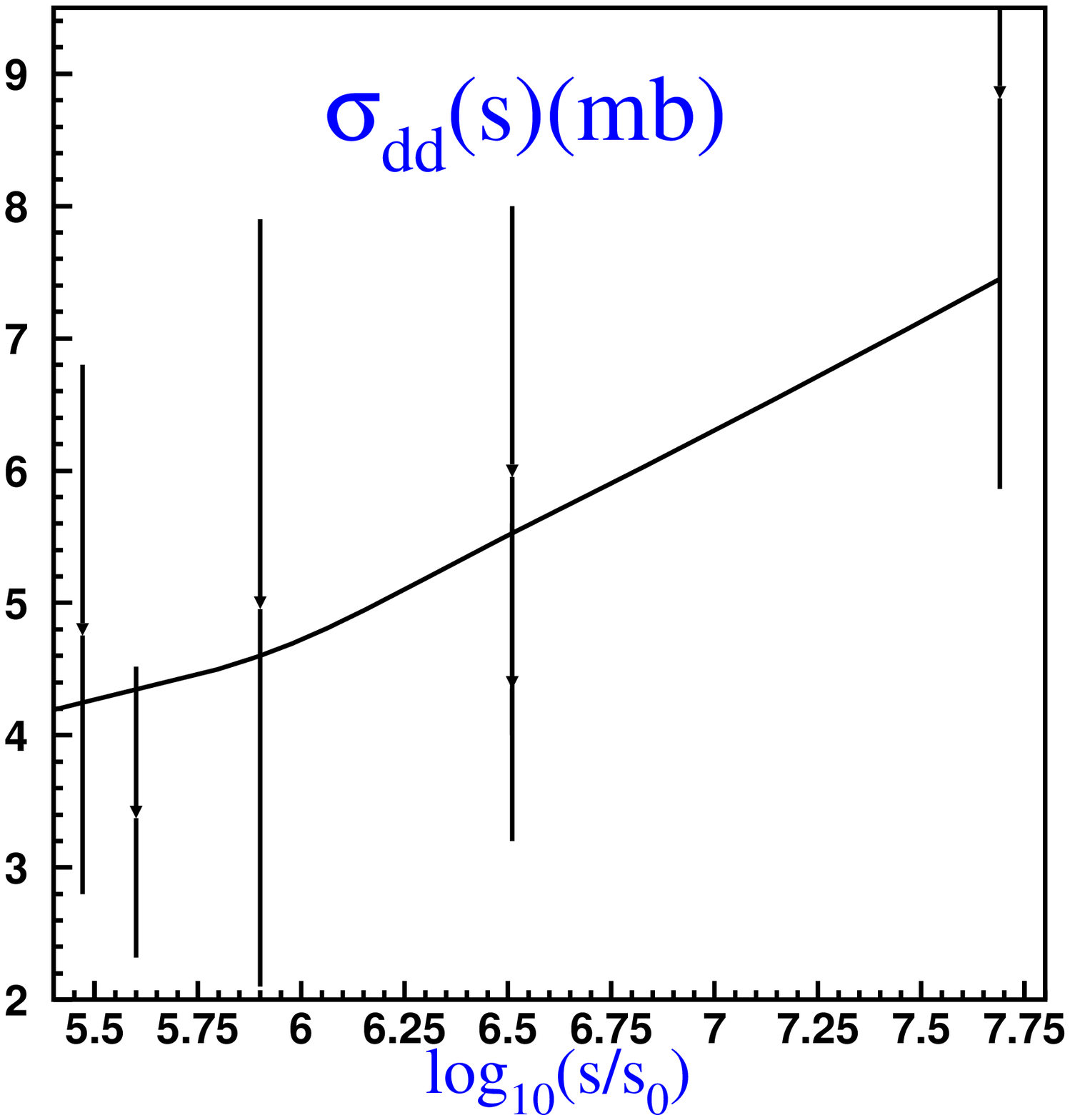,height=50mm,width=65mm}\\
\fig{fit}-c & \fig{fit}-d\\
\epsfig{file=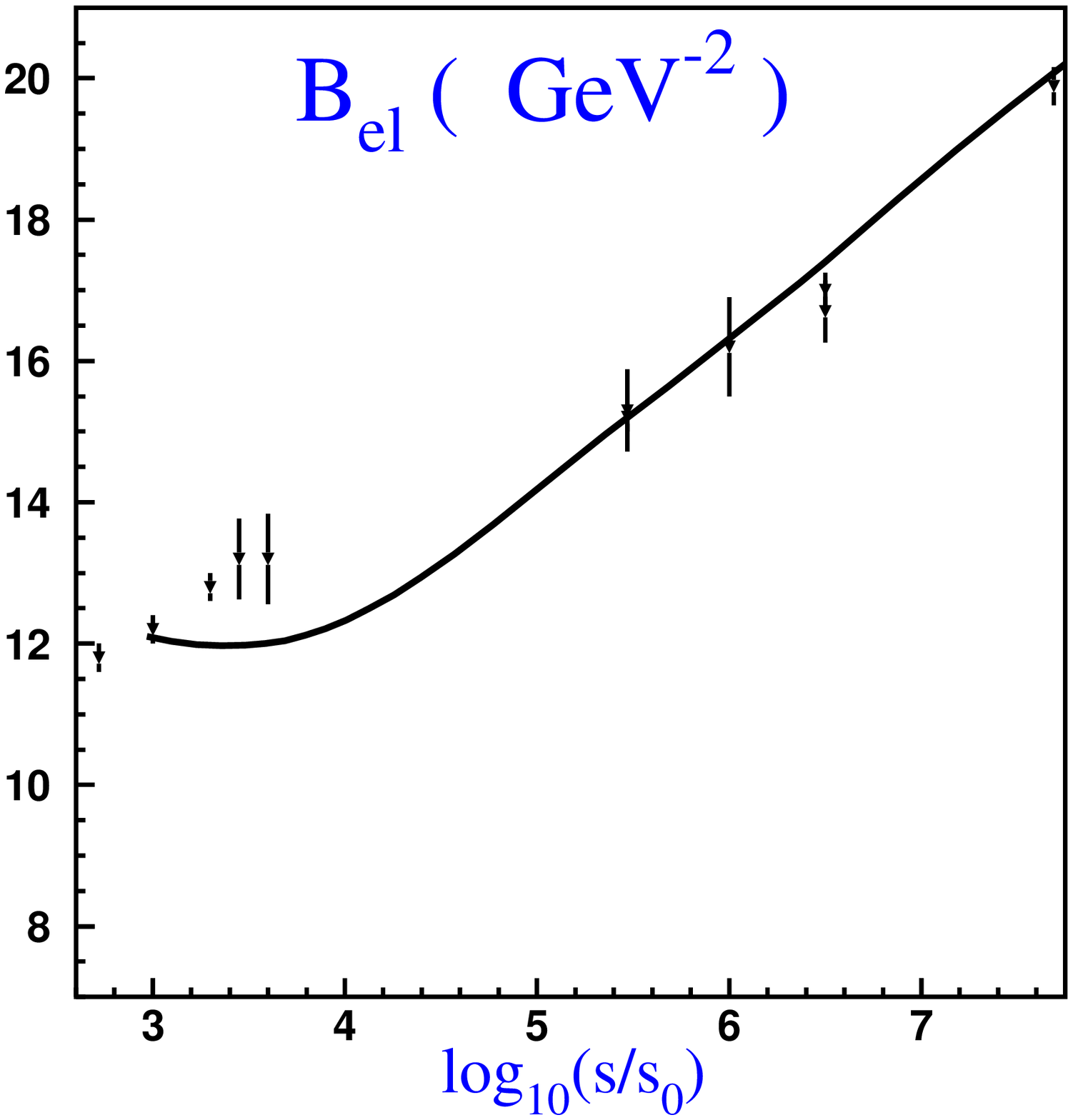,height=50mm,width=65mm}
 &\epsfig{file=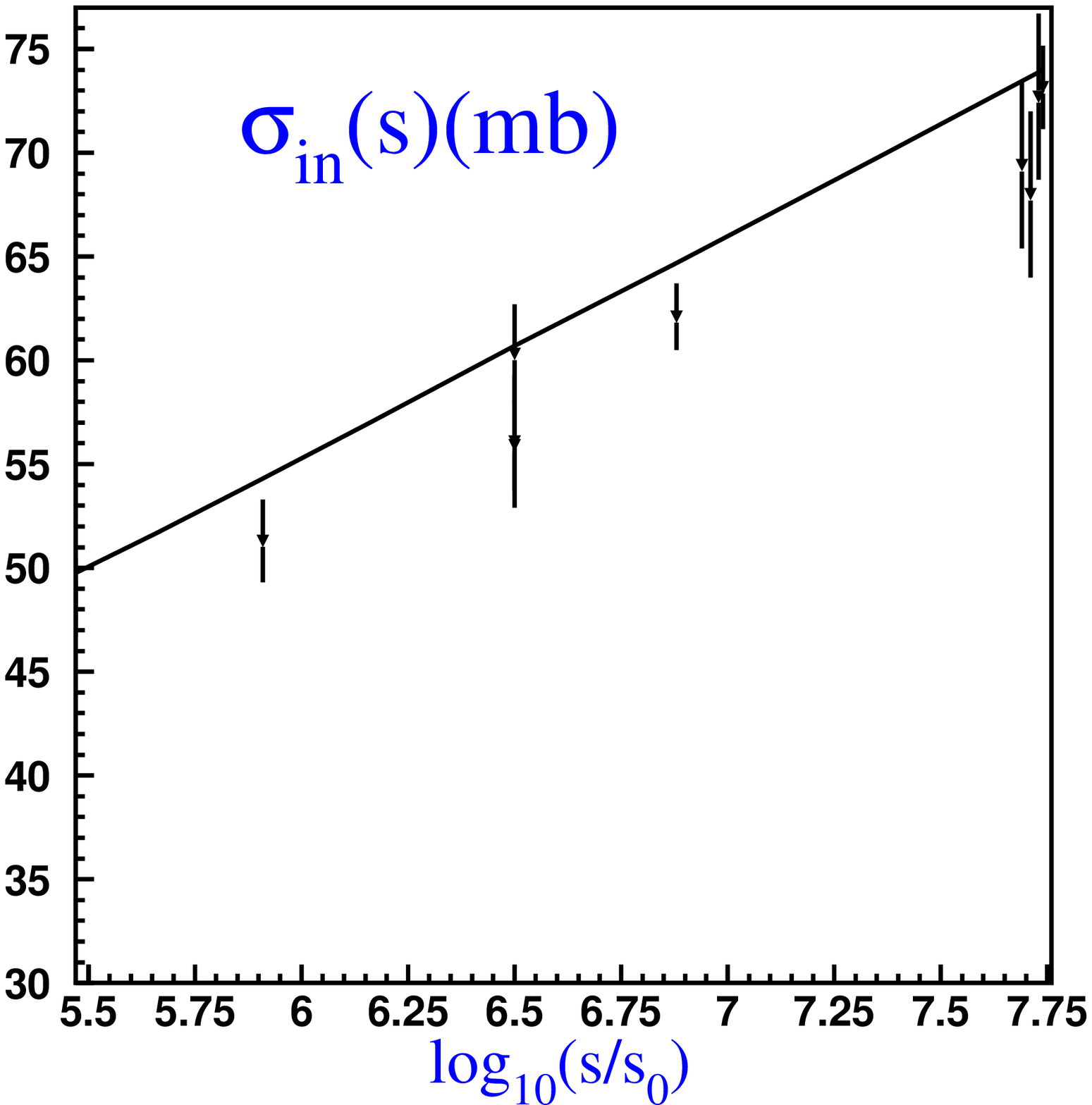,height=50mm,width=65mm}\\
\fig{fit}-e & \fig{fit}-f\\
\end{tabular}
\caption{Comparison with the experimental data the energy behaviour
 of the total (\protect\fig{fit}-a),
elastic  (\protect\fig{fit}-b), single diffraction  (\fig{fit}-c),
double diffraction (\protect\fig{fit}-d) and inelastic (\protect\fig{fit}-f) 
cross sections
and elastic slope( \protect\fig{fit}-e), 
The solid lines show our present fit . The data has been taken from
Ref.\protect\cite{PDG} 
for energies less than the LHC energy. At the LHC energy for total and
elastic cross 
section we use TOTEM data\protect\cite{TOTEM} and for single and double
 diffraction 
cross sections are taken from Ref.\protect\cite{ALICE}.
}
\label{fit}}
%%%%%%%%%%%%%%%%%%%%%%%%%%%%%%%%%%%%%%%%%%%%%%%%%%%%%%%%%
%%%%%%%%%%%%%%%%%%%%%%%%%%%%%%%%%%%%%%%%%%%%%%%%%%%%%%%%%%%%%%%%%%%%%%
%%%%%%%%%%%%%%%%%%%%%%%%%%%%%%%%%%%%%%%%%%%%%%%%
\FIGURE[t]{
\centerline{\epsfig{file=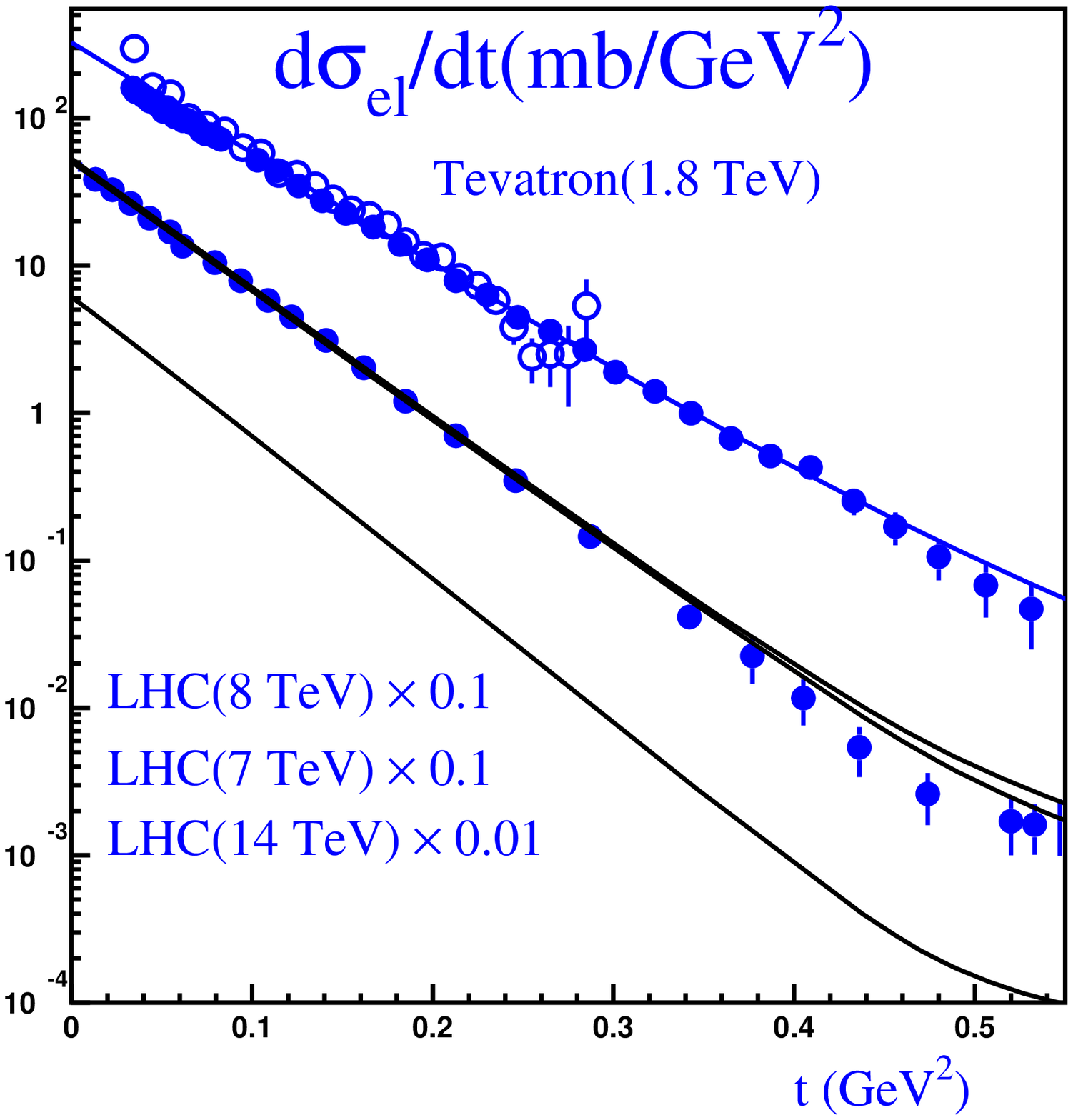,width=100mm,height=80mm}}
\caption{ $d \sigma_{el}/dt$     versus $|t|$ 
    at Tevatron (blue curve and data)) and LHC
 ( black curve and data) energies ($W = 1.8 \,TeV $ , $8 \,TeV$ , $ 7 \,Tev $
 and $ 14 \,TeV$ respectively).
 Data from Refs.\cite{E710,TOTEM}.}
\label{dsdt}}
%%%%%%%%%%%%%%%%%%%%%%%%%%%%%%%%%%%%%%%%%%%%%%%%
\section{Conclusions}
Our  goal in this paper is to deliver the  message: we
 have succeded in  building a model for soft interactions at high energy, 
which provides a good description all high energy data, including the LHC 
measurements. This
 model is based on the Pomeron with a
 large intercept ($\Delta_\pom = 0.23$) and slope $\alpha'_\pom \approx 
0$.
 We find no need to introduce two Pomerons: i.e. a soft and a hard one. 
The 
Pomeron in our model provides a natural matching with the hard Pomeron
 in  processes that occur at short distances.

The attractive feature of our model is that we are able to describe data 
over the whole energy range. Our description of 
 the  data at lower energies is slightly
 worse   than our description of
the data at high energy $W \geq 900\,GeV$. In addition we note that the
 qualitative features of the model are close to what we expect
 from N=4 SYM, which is the only 
theory that is able to treat the long distance physics on a
 solid  theoretical basis.

\section{Acknowledgements}
We thank all participants of ``Low x'2012 WS'' for fruiftul discussions on the subject.
This research of E.L.  was supported by the Fondecyt (Chile) grant 1100648. 
%%%%%%%%%%%%%%%%%%%%%%%%%%%%%%%%%%%%%%%%%%%%%%%%%%%%%%%%%%%%%%%%%%%%%%%%%%%%%%%%%%%

%%%%%%%%%%%%%%%%%%%%%%%%%%%%%%%%%%%%%%%%%%%%%%%%%%%

\begin{thebibliography}{99}

\bibitem{GLM1}
 E.~Gotsman, E.~Levin and U.~Maor,
  %``N=4 SYM and QCD motivated approach to soft interactions at high energies,''
  Eur.\ Phys.\ J.\  C {\bf 71} (2011) 1553
  [arXiv:1010.5323 [hep-ph]].
  
\bibitem{GLM2}
 E.~Gotsman, E.~Levin, U.~Maor and J.~S.~Miller,
  %``A QCD motivated model for soft interactions at high energies,''
  Eur.\ Phys.\ J.\  C {\bf 57} (2008) 689
  [arXiv:0805.2799 [hep-ph]].
  %%CITATION = EPHJA,C71,1553;%%
\bibitem{KAP}
 A.~B.~Kaidalov and M.~G.~Poghosyan,
  %``Description of soft diffraction in the framework of reggeon calculus.
  %Predictions for LHC,''
  arXiv:0909.5156 [hep-ph].
  %%CITATION = ARXIV:0909.5156;%%

\bibitem{KMR}
 A.~D.~Martin, M.~G.~Ryskin and V.~A.~Khoze,
  %``From hard to soft high-energy pp interactions,''
  arXiv:1110.1973 [hep-ph].
  %%CITATION = ARXIV:1110.1973;%%
\bibitem{OST}
S.~Ostapchenko,
  %``Monte Carlo treatment of hadronic interactions in enhanced Pomeron scheme:
  %I. QGSJET-II model,''
  Phys.\ Rev.\  D {\bf 83} (2011) 014018
  [arXiv:1010.1869 [hep-ph]].
  %%CITATION = PHRVA,D83,014018;%%
\bibitem{BST} 
 R.~C.~Brower, J.~Polchinski, M.~J.~Strassler and C.~I.~Tan,
  %``The Pomeron and Gauge/String Duality,''
  JHEP {\bf 0712} (2007) 005
  [arXiv:hep-th/0603115];\,\,
  %%CITATION = JHEPA,0712,005;%%
R. C. Brower, M. J. Strassler and C. I. Tan,
{\it "On The Pomeron at Large 't Hooft Coupling"}, arXiv:0710.4378 [hep-th].
 \bibitem{HIM}
Y.~Hatta, E.~Iancu and A.~H.~Mueller,
  %``Deep inelastic scattering at strong coupling from gauge/string duality :
  %the saturation line,''
  JHEP {\bf 0801} (2008) 026
  [arXiv:0710.2148 [hep-th]].
  %%CITATION = JHEPA,0801,026




 \bibitem{COCO}
 L.~Cornalba and M.~S.~Costa,
  %``Saturation in Deep Inelastic Scattering from AdS/CFT,''
 Phys. Rev. {\bf D 78}, (2008) 09010,
  arXiv:0804.1562 [hep-ph];\,\,\,
  L.~Cornalba, M.~S.~Costa and J.~Penedones,
  %``Eikonal Methods in AdS/CFT: BFKL Pomeron at Weak Coupling,''
  JHEP {\bf 0806} (2008) 048
  [arXiv:0801.3002 [hep-th]];\,\,
  %``Eikonal Approximation in AdS/CFT: Resumming the Gravitational Loop
  %Expansion,''
  JHEP {\bf 0709} (2007) 037
  [arXiv:0707.0120 [hep-th]].
  %%CITATION = JHEPA,0709,037;%%
  %%CITATION = JHEPA,0806,048;%%
    %%CITATION = ARXIV:0804.1562;%%

\bibitem{BEPI}
B.~Pire, C.~Roiesnel, L.~Szymanowski and S.~Wallon,
  %``On AdS/QCD correspondence and the partonic picture of deep inelastic
  %scattering,''
  Phys.\ Lett.\  B {\bf 670}, 84 (2008)
  [arXiv:0805.4346 [hep-ph]].
  %%CITATION = PHLTA,B670,84;%%
\bibitem{LMKS}
E.~Levin, J.~Miller, B.~Z.~Kopeliovich and I.~Schmidt,
 % {\it ``Glauber - Gribov approach for DIS on nuclei in N=4 SYM,''}
JHEP {\bf 0902} (2009) 048;\,\,
  arXiv:0811.3586 [hep-ph].
  %%CITATION = ARXIV:0811.3586;
\bibitem{ALICE}
M.~G.~Poghosyan,
  %``Diffraction dissociation in proton-proton collisions at $\sqrt{s}$ = 0.9 TeV, 2.76 TeV and 7 TeV with ALICE at the LHC,''
  J.\ Phys.\ G G {\bf 38}, 124044 (2011)
  [arXiv:1109.4510 [hep-ex]].
 ALICE ~Collaboration,
  {\it ``First proton--proton collisions at the LHC as observed with the ALICE
  detector: measurement of the charged particle pseudorapidity density at
  $\sqrt{s}$ = 900 GeV,''}
  arXiv:0911.5430 [hep-ex].
  %%CITATION = ARXIV:0911.5430;%%
  %%CITATION = ARXIV:1109.4510;%%
\bibitem{ATLAS}
G.~Aad {\it et al.}  [ATLAS Collaboration],
  %``Measurement of the Inelastic Proton-Proton Cross-Section at $\sqrt{s}=7$ TeV with the ATLAS Detector,''
  Nature Commun.\  {\bf 2} (2011) 463
  [arXiv:1104.0326 [hep-ex]].
  %%CITATION = ARXIV:1104.0326;%%
\bibitem{CMS}
CMS Physics Analysis Summary:
``Measurement of the inelastic pp cross section at √s = 7 TeV with the CMS detector", 2011/08/27.


\bibitem{TOTEM}
 F.~Ferro [TOTEM Collaboration],
  %``First data from TOTEM experiment at LHC,''
  AIP Conf.\ Proc.\  {\bf 1350} (2011) 172;\,\,\,G.~Antchev {\it et al.}  [TOTEM Collaboration],
  %``First measurements of the total proton-pro Abstract and Postscript and PDF from arXiv.orgton cross section at the LHC energy of \sqrt s =7TeV,''
  Europhys.\ Lett.\  {\bf 96} (2011) 21002,
  %``Proton-proton elastic scattering at the LHC energy of s** (1/2) = 7-TeV,''
  {\bf 95} (2011) 41001
  [arXiv:1110.1385 [hep-ex]].
  %%CITATION = ARXIV:1110.1385;%%
  %%CITATION = APCPC,1350,172;%%


  %%CITATION = EPHJA,C57,689;%%

\bibitem{GLMLAST}
  E.~Gotsman, E.~Levin and U.~Maor,
  %``Soft interaction model and the LHC data,''
  Phys.\ Rev.\ D {\bf 85} (2012) 094007
  [arXiv:1203.2419 [hep-ph]].
  %%CITATION = ARXIV:1203.2419;%%




\bibitem{GW}
M. L. Good and W. D. Walker,
Phys. Rev. {\bf 120} (1960) 1857.          

\bibitem{GU}
 G.~Gustafson,
  {\it ``The Relation between the Good-Walker and Triple-Regge Formalisms for Diffractive Excitation,''}
  arXiv:1206.1733 [hep-ph].
  %%CITATION = ARXIV:1206.1733;%%

\bibitem{MPSI}
A. H. Mueller and B. Patel:
{\it Nucl. Phys.} {\bf B425} (1994) 471;\\
A. H. Mueller and G. P. Salam:
{\it Nucl. Phys.} {\bf B475}, (1996) 293;\\
G. P. Salam:
{\it Nucl. Phys.} {\bf B461} (1996) 512;\\
E. Iancu and A. H. Mueller:
{\it Nucl. Phys.} {\bf A730} (2004) 460.
\bibitem{RY}
I. Gradstein and I. Ryzhik, 
{\it "Tables of Series, Products, and Integrals"}, Verlag MIR, Moskau,1981.


\bibitem{PDG}
C. Amsler et al. (Particle Data Group), Physics Letters, {\bf  B667}  (2008) 1. 

\bibitem{DL}
 A.~Donnachie and P.~V.~Landshoff,
  {\it ``Elastic Scattering at the LHC,''}
  arXiv:1112.2485 [hep-ph].
  %%CITATION = ARXIV:1112.2485;%%

\bibitem{BLHA}
M.~Block and F.~Halzen.
Phys. Rev. Lett. {\bf 107} 212002 (2011).

\bibitem{E710}
E710 Collaboration, Phys. Lett. {\bf B243} (1990) 158.\\
CDF Collaboration, Phys. Rev. {\bf D50} (1994) 5550.


\bibitem{RISTO}
R. Orava, Talk at Worshop:{\it `` Exclusive and diffractive processes in high energy proton-proton and nucleus-nucleus collisions"}, February 27 - March 2 , 2012, Trento, Italy;
http://diff2012-lhc.physi.uni-heidelberg.de/Talks/orava.pdf

\bibitem{JPR}
Jean Pierre Revol, {\it ``Proton-proton cross sections with multiplicites with ALICE at LHC''}, talk at ``Low X WS'', Paphos ,Cyprus, June 27 - July 1, 2012.

\end{thebibliography}
\end{document}